\documentclass[sigconf,nonacm]{acmart}
\usepackage{listings}
\usepackage{xcolor}
\setcopyright{none}
\AtBeginDocument{%
  \providecommand\BibTeX{{%
    \normalfont B\kern-0.5em{\scshape i\kern-0.25em b}\kern-0.8em\TeX}}}

\begin{document}
\title{How is Testing Related to Single Statement Bugs?}

\author{Habibur Rahman}
\email{habibur@ualberta.ca}
\authornote{Both authors contributed equally to this research.}
\authornote{This research was conducted while the author was a student at the University of Alberta.}
\affiliation{%
	\institution{University of Alberta}
	\city{Edmonton}
	\state{AB}
	\country{CA}
}

\author{Saqib Ameen}
\email{saqib1@ualberta.ca}
\authornotemark[1]
\authornotemark[2]
\affiliation{%
	\institution{University of Alberta}
	\city{Edmonton}
	\state{AB}
	\country{CA}
}

\begin{abstract}
	In this study, we analyzed the correlation between unit test coverage and the occurrence of Single Statement Bugs (SSBs) in open-source Java projects. We analyzed data from the top 100 Maven-based projects on GitHub, which includes 7824 SSBs. Our preliminary findings suggest a weak to moderate correlation, indicating that increased test coverage is somewhat reduce the occurrence of SSBs. However, this relationship is not very strong, emphasizing the need for better tests. Our study contributes to the ongoing discussion on enhancing software quality and provides a basis for future research into effective testing practices aimed at mitigating SSBs.
	\end{abstract}	


\maketitle
\settopmatter{printfolios=true}
\section{Introduction}
Writing unit tests is a common practice in the industry to ensure software quality. Usually meeting certain criteria of test coverage serves as a segway to product release. This is essential because software systems are used in critical areas such as health, security, finance, and space missions. In such settings, a faulty system is not acceptable. International Organization for Standardization (ISO) also mandates testing as a part of the software development life cycle. Despite all efforts, the software system may remain prone to bugs. This raises a question on the effectiveness of testing. Is it even useful?

\begin{figure}[h]
    \centering
    \includegraphics[width=1.0\linewidth]{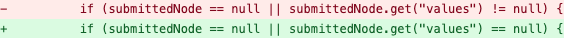}
    \caption{Example of a single statement bug before and after the fix.}
    \label{fig:ssb}
\end{figure}

Multiple studies tried to answer this question under different settings but there seems to be no consensus on it. These studies examined all types of bugs that were found in the systems. Nobody studied the effectiveness of testing on subsets of the bugs found in the software systems. One important subset of bugs is known as Single Statement Bugs (SSBs). They appear in just a single statement and can be fixed by modifying that statement. Those modifications can be as simple as changing a variable name, ordering arguments in a function, changing the return type, and so on. Figure \ref{fig:ssb} shows an example of a single statement bug where $!=$ needs to be replaced with $==$.

\begin{figure}[h]
	\centering
	\includegraphics[width=\linewidth]{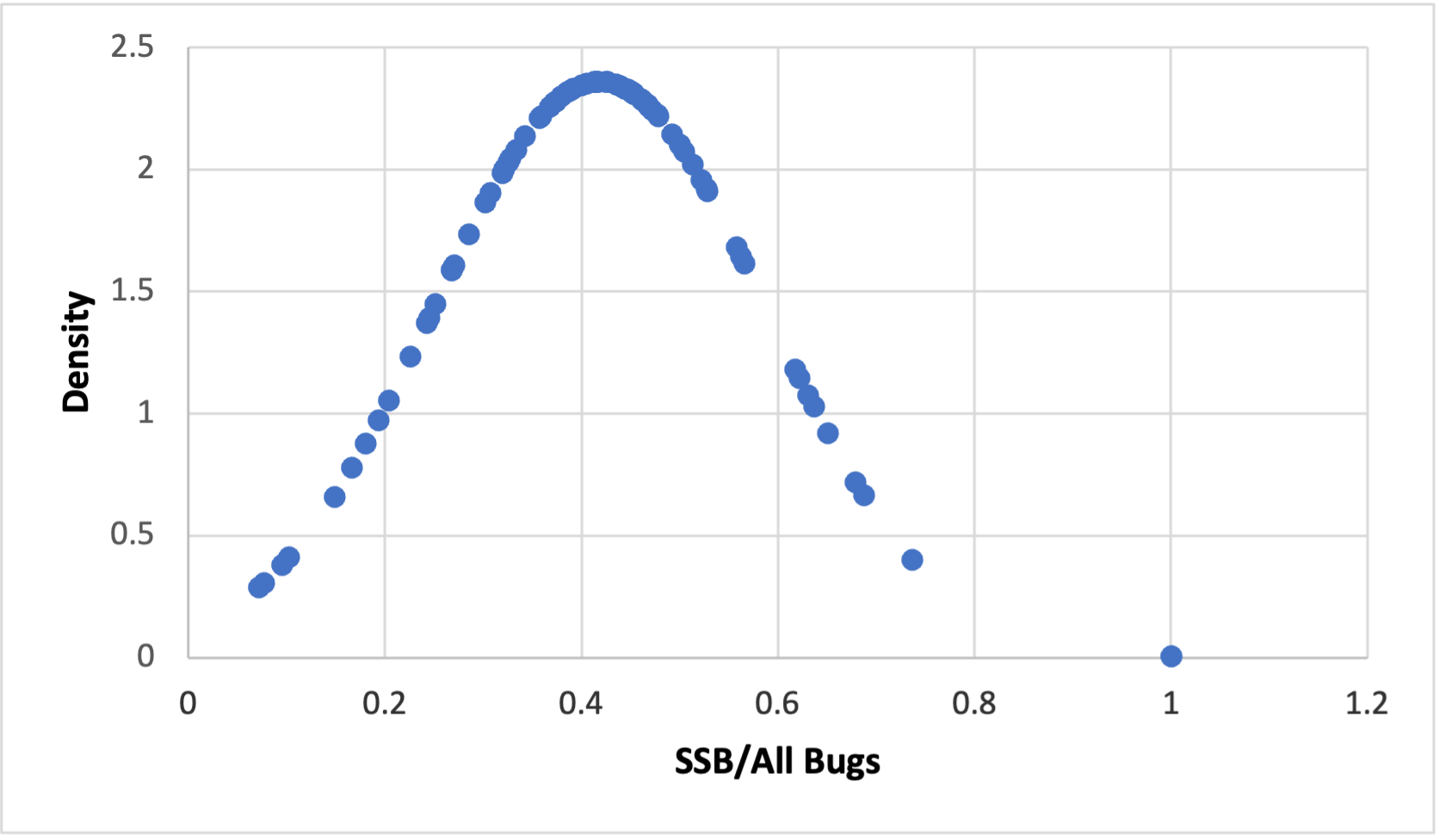}
	\caption{Distribution of SSBs in the top 100 open source Java projects on GitHub using Maven build system}
	\Description{A graph of density of single statement bugs.}
	\label{fig:density}
\end{figure}

SSBs occur quite often \cite{sstubs} and can be very critical. For example, Apple’s return bug resulted in an invalid SSL/TLS connection verification, putting the sensitive data of millions at risk. The normal distribution of the ratio of SSB/all bugs for top 100 open source java projects using the Maven build system on GitHub has a bell curve shown in Figure \ref{fig:density}. We can see that the mean lies around > 0.4. This means there are more than 40\% of SSBs in those projects. Mitigating this large chunk of bugs can be useful in improving the overall quality of the software and prevent failures in production. The goal of our study to help practitioners understand the effectiveness of testing for SSBs.

We intend to understand the effectiveness of testing on SSBs by the correlation between test coverage and single statement bugs. If more test coverage in a project results in fewer SSBs, it shows that testing is helpful and vice versa. This leads to our research question, \emph{Is there a correlation between test coverage and single statement bugs?} We hypothesized that there is a weak to moderate relationship between SSBs and test coverage, i.e., testing is helpful to some extent.

\begin{figure}[h]
	\centering
	\includegraphics[scale=0.7, angle=-90,width=0.5\linewidth]{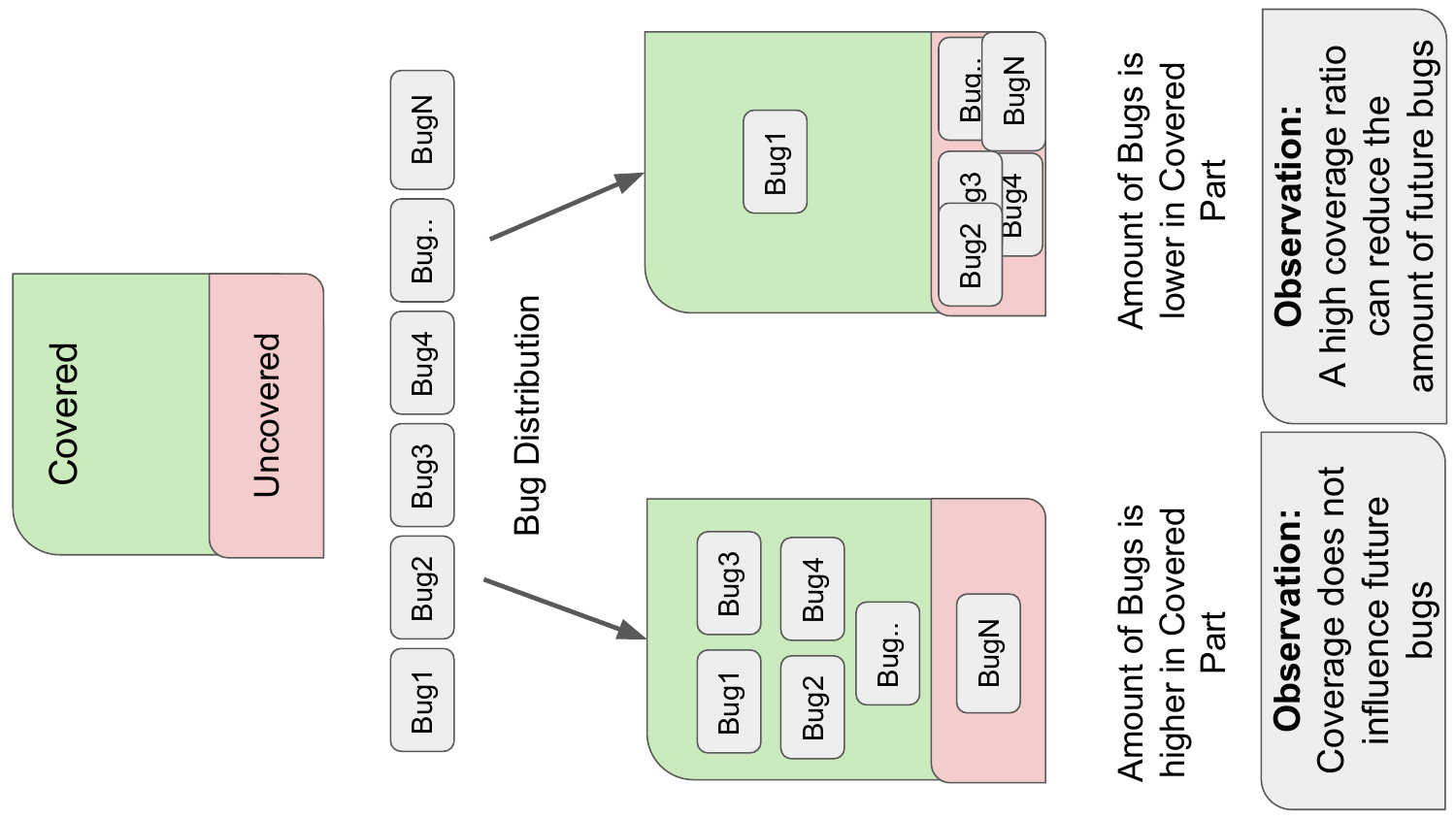}
	\caption{An overview of the bug distribution scenarios based on coverage. \cite{similar}}
	\Description{A snap of the bug distribution scenarios based on coverage.}
	\label{fig:cov_scn}
\end{figure}

To verify our hypothesis, we analyze the post-unit test bugs in the areas covered and not covered by the testing. If a part is covered by testing, it means it has been executed for its intended purpose and it works fine. In future, if more bugs are found in the covered part, it will show that testing did not help here. However, if that is not the scenario, it indicates the effectiveness of testing. This approach is similar to Bach et al.\cite{similar} as shown in Figure \ref{fig:cov_scn}. To check that, we first generate the coverage report at the release time, and see if the bugs found after the release are in the covered part or not covered part. Finally, we calculate the percentage of coverage and percentage bugs in the not covered part, to find the correlation between them. It serves as a proxy for the effectiveness of unit testing for SSBs.

For our study, we use the Mining Software Repositories (MSR) 2021 challenge' SSBs dataset\cite{sstubs}. It contains 7824 SSBs from the top 100 Maven-based open source Java projects on GitHub. The important thing about this dataset is that projects can be built and their test can be executed. Which is important for us to be able to generate reports.

After the experiment, we have found that there is a weak to moderate correlation between the test coverage and SSBs. This shows that testing is effective to some extent in reducing the SSBs.

\section{Background and Terminology}
Test coverage is the percentage of lines of code executed by the tests for a project. We measure testing in the form of test coverage. A project that has 87\% coverage means 87 out of 100 lines of code has been executed by the test cases of that project. In this writing, the word 'coverage' refers to test coverage.

In statistics, correlation is the statistical relationship between two random variables or bivariate data. Correlation is measured in terms of the correlation coefficient that ranges from -1 to 1 where -1 means negative correlation, 0 means no correlation, and 1 means a positive correlation between two variables. The correlation coefficient is useful for hypothesis testing. Based on the value of the correlation coefficient, a hypothesis can be accepted or rejected. There are several methods to calculate this coefficient, Pearson correlation coefficient is one of them. It can be expressed as the equation below:
\begin{equation}
r_{xy} = \dfrac{\sum{(x_{i}-\bar{x})(y_{i}-\bar{y})}}{\sqrt{\sum{{(x_{i}-\bar{x})}^{2}{(y_{i}-\bar{y})}^{2}}}}
\label{eq:1}
\end{equation}
Where:

$r_{xy}$ – the correlation coefficient of the linear relationship between the variables x and y

$x_{i}$ – the values of the x-variable in a sample

$\bar{x}$– the mean of the values of the x-variable

$y_{i}$ – the values of the y-variable in a sample

$\bar{y}$ – the mean of the values of the y-variable

Percentage test coverage, and number of bugs in the not covered part are the variable for our case.

\section{Related Work}

Our work overlaps with two niches of the software engineering literature. The first one is related to analyzing the relation of test coverage and its effectiveness. They utilize different techniques, under different conditions to study the usefulness of test coverage. The second one is related to the usage of SSB for an empirical study or a software application.

In the first kind of work, more or less, researchers aim to find the correlation between test coverage and the occurrence of bugs. To date, no consensus exists in the community on the usefulness of test coverage. Some studies tend to agree, while others disagree. For example, Gren and Antinyan \cite{gren_2017} and Antinyan et al.  \cite{ericson} found a weak to no correlation between the test coverage and post-unit test defects in their separate studies and concluded that test coverage was not helpful. Both of them conducted the study on a single, but large-scale industrial project. The former one did not reveal the details of the project, while the latter one worked with Ericson. Ericson mainly deals with telecommunications and networking software around the world. One aspect of these studies which is similar to our work is that they rely on actual bugs found after unit tests and real test cases written by developers. In contrast to our case, both of them considered all types of bugs, not just the SSBs. The methodology of Gren and Antinyan \cite{gren_2017} also differs from our approach, they evaluate coverage in terms of files and only consider the files with either 100\% coverage or no coverage at all. While Antinyan et al. \cite{ericson} considered the overall coverage, which is similar to our approach. Due to the small sample size and specific software niche, the results of both studies cannot be generalized for other studies.

On the other end of the spectrum, Inozemtseva and Holmes \cite{waterloo} found a weak to moderate correlation between the test coverage and its effectiveness, when the test suite size was controlled for. They conducted the study on five large open-source Java projects powered by the Ant build system and during the study. Contrary to our study, they used synthetic test suits and mutation testing to generate faulty programs. While they are a good approximation, they do not necessarily reflect the real scenarios and are limited by the algorithms behind those tools. Furthermore, they considered an extra variable, i.e., the test suite size, and varied it throughout the study while generating test cases. In our approach, we do not rely on any tests or bug generators. In another study, Namin and Andrews \cite{waterloo2} also considered the role of test suite size in addition to the coverage on the effectiveness of unit tests. They found that by increasing the coverage, indirectly, the test suite size is increased, which increases the effectiveness of tests. For varying, but controlled test suite sizes, they found a high to a weak correlation between the coverage and test effectiveness. Similar to the aforementioned study, they also relied on self-generated test cases and considered only the Siemen suite of seven (C, C++) programs. Our technique is similar to what Bach et al. \cite{similar} have used in their study related to the impact of coverage on bug density. But we use a simpler approach since we only have SSBs. In comparison to our study, they considered all types of bugs and analyzed a single project, i.e., SAP HANA, and found a positive impact of coverage on bugs density. In general, the literature work on this topic differs from our work in one or more of the following areas:
\begin{itemize}
	\item A very small number of projects were considered.
	\item Artificially generated test cases were used.
	\item Synthetic bugs were introduced in the system.
	\item All types of bugs were considered.
\end{itemize}

Furthermore, in studies, done in collaboration with the industry, the details of the analyzed software were not released. Which is a barrier in generalizing or understanding the results from those studies. There could be additional internal factors and software engineering practices leading to those results. Lastly, our dataset differs from those used in the aforementioned studies.

The second part of the literature, which uses SSB mainly differs from our work in terms of intended implicatications. To the best of our knowledge, at this point, the use of SSB in literature is limited to program repair. There is no study related to test coverage and its effectiveness. Chen et al. \cite{sequenceR} used a combination of Bugs2Fix \cite{bugs2fix} and CodRep \cite{coderep} single statement dataset to propose a novel learning-based program repair technique. Their dataset is also different from the one we are using. Their dataset is not intended for building projects or running tests as there is no information on the project's build system or even if they can be run or not. Those factors are important for us to be able to generate the coverage reports. Another study by Karampatsis and Sutton \cite{sstubs} also uses SSBs. In their study, they provided a new dataset on SSBs and all types of bugs in the top 100 open-source Java projects on GitHub and attempted to find the frequency of occurrence of SSBs. They found out that SSBs occur with a frequency of one bug per 1600-2500 lines of code. Even more important contribution of this paper is that the projects in their dataset use the Maven build system and can be built. If there are tests in the project, they can be executed. We use their dataset of SSB and real test cases present in those projects to conduct this study on the effectiveness of coverage on SSBs.

\section{Methodology}

This section describes the detailed methodology used to process the information from the dataset and curates the desired data for evaluation. It starts with the description of the dataset for a better context and then goes through each step describing its purpose, input, and output.

\begin{figure}[h]
	\centering
	\includegraphics[width=1.0\linewidth]{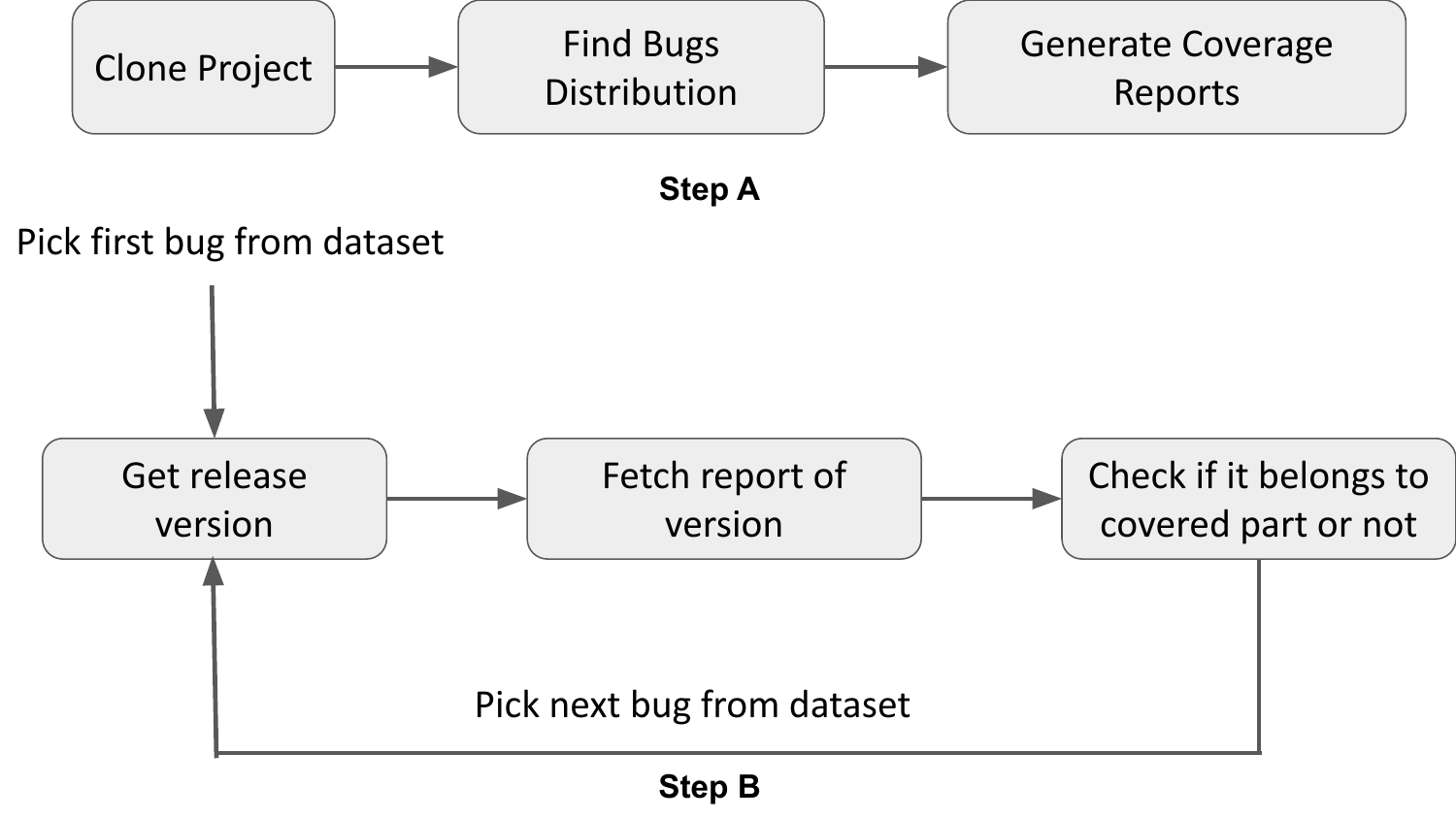}
	\caption{An overview of the complete methodology.}
	\Description{A snap of overall methodology.}
	\label{fig:method_snap}
\end{figure}

Our main goal is to find out if there are more bugs or fewer bugs in the covered part, after unit tests were written. For this purpose, an overview of our methodology is shown in Figure \ref{fig:method_snap}. It consists of two major steps. In step (A), we generate the coverage reports. In the second step (B), with the help of those coverage reports, we try to find the is a certain bug belongs to the covered part or not. At the end of this whole process, we have percentage coverage for each project, and the percentage covered of SSBs in the covered as well as uncovered parts. We prune the resulting data for any outliers and then use Pearson's correlation to find our results.

\subsection{Dataset}
For our study, we used the ManySStuBs4J\cite{sstubs} dataset. This corpus consists of simple fixes to SSBs found in top open source Java projects on GitHub. It has two different variants. The first one contains teh top 100 open source Java projects which use Maven as a build system. They can be built, and tests in those projects can be executed. The second one contains the top 1000 open source Java projects. In this larger variant, the projects use different build systems and they may or may not be built. For our study, a natural choice is to use the former variant due to testability. Table \ref{tab:stat}, shows the dataset statistics.

\begin{table}[h]
	\begin{tabular}{|c|c|c|c|c|}
		\hline
		\textbf{Projects}&\textbf{Bug}&\textbf{Buggy}&\textbf{Bugs}&\textbf{SSBs}\\
		& \textbf{Commits}&  \textbf{stmts} & \textbf{per}&\\
		& & & \textbf{Commit}&\\
		\hline
		100 Java Maven&12598		   & 25539	&	          2.03    &  		7824\\
		1000 Java&86771   	   & 153652	&	          1.77     &  		51537\\
		\hline
	\end{tabular}
	\caption{Statistics of the ManySStuBs4J dataset.}
	\label{tab:stat}
\end{table}
Figure \ref{fig:database_snap}, shows an example of the ManySStuBs4J dataset instance. In this dataset snapshot, we can see that it contains a bug type, project name, fix commit SHA, parent commit SHA, Bug Line number, and other properties. For our study, we have used the project name, fix commit SHA, parent commit SHA, and bug line numbers from each of the data items. In the $projectName$ property, the part before period represents the organization name or user name and the part after the period represents the name of project. It comes in handy while cloning the repos for processing in the later sections. For the sake of brevity, we omitted some parts in the snapshot, which are represented by three dots.

\begin{figure}[h]
	\centering
	\begingroup
	\fontsize{7pt}{10pt}\selectfont
	\begin{verbatim}
{
"bugType": "CHANGE_OPERATOR",
"fixCommitSHA1": "aa90e04b5e6eb7f6d46dde16867196329568324e",
"fixCommitParentSHA1": "46d3a4007fe1418d53baabc16dec39275079684b",
"bugFilePath": "/java/org/../GetRuntimeFormDefinitionCmd.java",
"fixPatch": "...",
"projectName": "Activiti.Activiti",
"bugLineNum": 184,
"bugNodeStartChar": 8444,
"bugNodeLength": 35,
"fixLineNum": 184,
"fixNodeStartChar": 8444,
"fixNodeLength": 35,
"sourceBeforeFix": "submittedNode.get(\"values\") != null",
"sourceAfterFix": "submittedNode.get(\"values\") == null"
}
	\end{verbatim}
	\endgroup
	\caption{A snapshot of the ManySStuBs4J dataset instance.}
	\Description{A snapshot of the ManySStuBs4J dataset instance.}
	\label{fig:database_snap}
\end{figure}

\subsection{Splitting the Dataset By Projects}

The corpus of the dataset is quite large. While processing a single project, we should not be enumerating through the whole dataset. To ease the search, we split the dataset based on the project name. We processed the whole dataset to sort it by the project name and then saved data of each project in separate JSON files. In each of the new datasets, they contain bugs from the same project.

\subsection{Cloning Projects}

To clone the projects, first we constructed the repository link for each project by using the $projectName$ property given in the dataset. Then we used those links to clone the projects from GitHub in our local machines for further processing. The whole process was automated through a Python script.

\subsection{Finding the Distribution of Bugs}

This is a crucial step. The bugs listed in the ManySStuBs4J\cite{sstubs} dataset span multiple years. For example, in the case of JUnit4, there are bugs from the year 2002 till 2017. To check, if a bug was discovered in the covered portion or not, we need to get the coverage report at the release time before that bug was found. Since we target a certain coverage before each release, the release time serves as a good point to take the coverage report. But it is very difficult and time-consuming to generate reports for such a long time for various reasons. Firstly, it is hard to resolve the dependencies to execute the test cases, and there could be a very large number of releases in such a long period. Secondly, for the very old versions, it hard to integrate Jacoco to begin with. Jacoco is the tool that we use to generate the coverage reports.

\begin{figure}[h]
	\centering
	\includegraphics[width=0.6\linewidth]{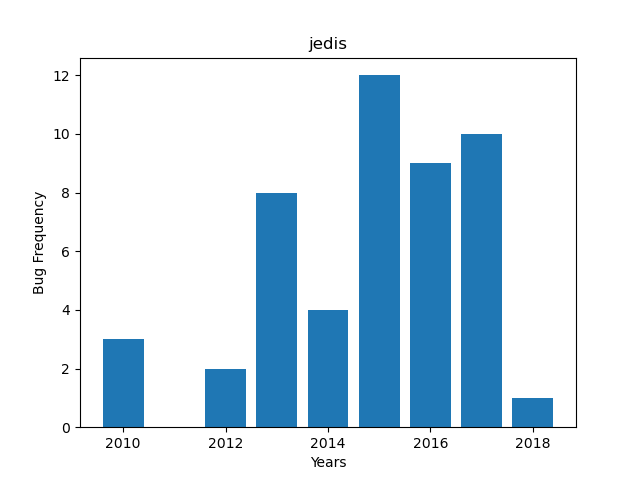}
	\caption{Distribution of SSBs in project Jedis}
	\Description{A graph of frequency of SSBs over the years}
	\label{fig:distribution}
\end{figure}

To resolve this, we find the distribution of the bugs with respect to years. Then we identify the high bug density area to generate the reports for that portion. It serves as a good proxy for the overall project. For example, Figure \ref{fig:distribution} shows an example of such distribution in the form of a bar graph for a project in the dataset, named Jedis. It is clear from the image that it is more useful to generate the reports for the releases between 2013 and 2017 than any other portion.

For some projects, it was not possible to generate the reports for all the identified regions for a few reasons. For some versions the dependencies could not be resolved, for a few jar files were missing, some used very old versions of Maven, and in a few cases, there were no tests. Some projects had very huge test suites, taking more than an hour to execute the tests and generate the reports. In such cases, it was not possible to generate reports for a good number of versions. Another advantage of this approach is that we end up generating coverage reports for the consecutive releases. In consecutive releases, coverage does not change much. As a result, when we can take the average of coverage to indicate the percentage coverage in the project.

\subsection{Generating Coverage Reports}

Once we identified the versions using the bug density distribution, we manually checked out each version and resolved the dependencies to generate the coverage reports. Most of the projects did not include Jacoco to generate the reports. In such projects, we had to figure out the appropriate version for each release before generating the report. Each project also has a different structure and test runner to execute the unit tests. That also needed to be taken into account to be able to generate the reports.

We evaluated more than 50 projects from the dataset during our study. However, we were able to generate reports properly for around 17 projects only. It is mainly due to failure in resolving one or more issues mentioned above. All the generated reports and files with fixes for each project are available as a part of the recreation package for this study in the GitHub repo of the project artifacts.

\subsection{Getting Release Number}

Once we have all the reports for the projects. The next step (B) of our methodology begins, where we have to analyze if find the ratio of bugs in the covered and uncovered parts. This whole process is automated with Python and Shell scripts.

For this purpose, we used the separated dataset of projects. For each project, we enumerated through its dataset, going through each SSB one by one. For each SSB, we identified its release version and then used the report generated for that version to identify it that bug was in the covered part or not. This process is repeated for all the bugs in the dataset against that project and the results are recorded.

\subsection{Fetching Report for Particular Release}

At this stage, we fetch those reports based on the release version. During the manual report generation process, we saved the $xml$ file as well as the $html$ file for the Jacoco report. The former contains the details about the covered and uncovered lines at a granular level while the latter contains the percentage coverage for the test suite.

We parsed the $xml$ file using the MiniDom Python API to identify in which part the bug occurred and then parsed the $html$ file with BeautifulSoup (an open source library for scrapping) to extract the percentage coverage for that particular version.

\subsection{Identifying SSBs in Covered and Not Covered Parts}

In this section, we explain how exactly we identified and counted the number of bugs in covered and not covered parts from the $jacoco.xml$. The $xml$ report file contains information like missed instructions, missed branches, as well as covered instructions and covered branches. For the bugs, where the covered instructions and covered branches were zero, they corresponded to the uncovered part and vice versa.

After enumerating the whole dataset, and repeating the aforementioned step for each of them, we end up with the total number of bugs in the covered and not covered part. Using those numbers, we find the percentage of bugs in the uncovered part.

\subsection{Calculating the Correlation Coefficient}

The data collected till this stage is shown in Table 2. It does not reflect all the data from section 4.8, rather it represents the data after removing the outliers. Outliers projects included a very small number of bugs (less than 5) and all of them were either in the covered part or the uncovered part. The reason such outliers occurred, despite identifying the high bug density releases in section 4.1, is that we could not generate reports for many versions in those parts.

We used this data to calculate the correlation coefficient between the percentage of SSBs that are not covered part and the average percentage coverage. For calculating correlation, we used the Equation \ref{eq:1}. Based on this correlation coefficient, we verified the hypothesis. The results and their implications, based on the result of this step, are discussed in detail in Section 5.

\section{Results}
Previous section explains how we used the SStuBs to collect the data from GitHub, used Jacoco to generate the coverage reports, and processed them to get the desired data. In this section, using that data, we try to answer our research question.

\subsection{How is Testing Related to SSBs?}

\begin{figure}[h]
	\centering
	\includegraphics[width=\linewidth]{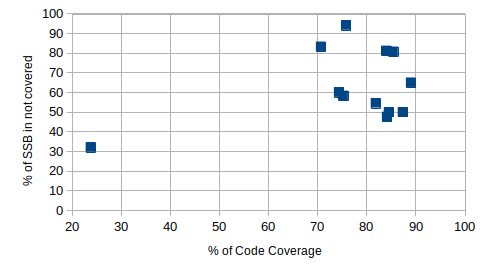}
	\caption{Correlation of the percentage of bugs in not covered part and the average of percentage test coverage}
	\label{fig:corr}
\end{figure}

In our research question, we asked if there a correlation between test coverage and SSBs. The purpose of this question is to find out if increasing the coverage can be “actually” helpful in reducing the SSBs.
\begin{table}[h]
	\begin{tabular}{|c|c|c|}
		\hline
		\textbf{Project Name} & \textbf{Percentage} & \textbf{Bugs in Not} \\
		&\textbf{Coverage}& \textbf{Covered (\%)}\\
		\hline
		alibaba.druid &75.25&58.33\\
		alibaba.fastjson & 87.33&	50\\
		AsyncHttpClient.async-http-client &74.33&60\\
		brettwooldridge.HikariCP &75.78&	94.12\\
		Bukkit.Bukkit &23.75&32\\
		cucumber.cucumber-jvm &84.5&50\\
		google.auto &85.43&80.769\\
		google.closure-compiler &84.11&	47.62\\
		google.guice&77&65\\
		jhy.jsoup&81.83&54.55\\
		junit-team.junit &84&81.25\\
		mybatis.mybatis-3 &70.67&83.33\\
		\hline
	\end{tabular}
	\caption{Data collected about projects on percentage coverage and percentage of SSBs in not covered part}
	\label{tab:ppc}
\end{table}
Table \ref{tab:ppc}  shows the data collected to answer this question. The first column shows the project name, and the second column shows the average percentage test coverage, while the third column shows the percentage of SSBs in the not covered part. We are considering the bugs in the non-covered part as they provide an easy way to understand the effectiveness of coverage. A higher percentage of them indicates that coverage is effective and vice versa. In our data, we have a mixed percentage. We used this data to find the correlation coefficient (\emph{r}) between the percentage of coverage and the percentage of SSBs. The \emph{r}-value turned out to be 0.40, which translates into a weak to moderate positive correlation between them. It shows that high coverage helps mitigate the SSBs. This correlation is shown in \ref{fig:corr}. We can see that increased covered is related to greater percentage of bugs in the not covered part and testing is useful here.\\

\fbox{\begin{minipage}{25em}
		\textbf{RQ 1 Answer.} \emph{Our results suggest that there is a positive weak to moderate correlation between the unit test coverage and the number of SSBs found in the not covered parts. The coverage seems to be effective for SSBs, to some degree.}
	\end{minipage}}\\\\

Our study has a few important implications. Firstly, it is one of its kind to explore the effectiveness of coverage on the SSBs and the results indicate that it’s a promising research area to further investigate. Further studies will help better understand the nature of this relationship under different settings and we might be able to reach a consensus about the unit test effectiveness. Secondly, knowing that testing is helpful, will help the practitioners allocate resources and prioritize testing more effectively. Which in turn, can help improve the software quality.

\section{Threats to the validity}
In this section, we discuss some of the possible threats to the validity of our study. As described in section 4, we considered multiple versions of each project and wrote our scripts for processing and some tasks automation. This involvement of the human factor equates to the possibility of human error in the process. There could be a mismatch in the versions, and we could have made some unintentional mistakes in the code, however, throughout the process we also did manual verification to reduce this risk. The results of our study are also limited by the SStuBs(sstubs) dataset we used. While it includes a good number and range of projects from different domains, it is limited by the language type and build system. All the projects use the Maven build system and are developed in Java. The results obtained from the study might not apply to the other type system, languages, or even other build systems.

Another important aspect to consider is that all the reports are generated manually by checking out each version one by one, resolving dependencies, and configuring Jacoco. It is not only a tedious manual process but also makes it highly susceptible to error. However, throughout the process, we did manual verification to mitigate this threat as much as possible. Another threat is that the projects included in the dataset are the top open-source Java projects. They are maintained for more than a decade by the open-source community and used by tons of organizations and developers. Through their feedback and community involvement, they have matured over the years. Since we only consider the high-density bugs areas where we have tests written and we can generate reports, they might not reflect their whole project life cycle. Lastly, since we only considered open source projects, our results might not apply to the close source projects.

\section{Conclusion and Future Work}

We tried to find the effectiveness of the unit testing for the SSBs. Instead of synthetic tests or faults, we relied on the real bugs found in the software systems and the real tests written at the time of their releases. We used a good number of projects so that the result could be generalized for the Maven based Java projects. All the projects are open source and can be explored to understand their nature. We found that to some extent, fewer bugs were found in the covered parts as compared to the uncovered parts. Which shows that coverage should be improved in software systems to reduce the number of SSBs. For the companies, it implies that testing should be prioritized since it is helps ensure the software quality.

Apart from the aforementioned implications, our study paves the way for potential future work. We identified a few directions which can be explored in the future to better understand the effect of coverage on the SSBs. We identified in related work (Section 4) that some studies added more variables, such as test suite size, while studying the effectiveness of testing. A similar approach can be taken with SSBs to see if the correlation still holds or changes. Secondly, the scope of our study is limited to only one build system, i.e, Maven, and Java language. This is mainly due to the availability of buildable and testable softwares in Maven along with the SSBs dataset. Similar studies can be conducted with other build systems or even other languages. Lastly, all the projects included in our study were open source projects. It would be interesting to see the results in closed source projects.

The artifacts of this project are available in our GitHub repo at the following link:
\href{https://github.com/habibrahmanbd/SSBandTesting}{https://github.com/habibrahmanbd/SSBandTesting}

\begin{acks}
	We would like to thank Sarah Nadi for her guidance and teaching us the necessary skills for this project. We would also like to acknowledge the CMPUT501 TAs, Batyr and Henry, for their assistance during lab sessions and helping us understand different technologies. Additionally, we note that this research was conducted while we were students at the University of Alberta.
\end{acks}

\bibliographystyle{ACM-Reference-Format}
\bibliography{sample-base}

\appendix

%
%
%
%
%
%
%

\end{document}